\documentclass[pal,aps,twocolumn,amsmath,amssymb]{revtex4-1}
\usepackage{epsfig}
\usepackage{amssymb,amsmath}
\usepackage{color}
\usepackage{graphicx}

\newcommand{\bhat}[4]{\hat{b}_{#1 #2 #3}^{#4}}
\newcommand{\nhat}[3]{\hat{n}_{#3 #1 #2 }}

\newcommand{\hc}{\mathrm{h.c.}}
\newcommand{\xvec}{\mathrm{\bf r}}

\newcommand{\BigO}[1]{\ensuremath{\mathcal{O}\left(#1\right)}}

\begin{document}
\title{Vortex macroscopic superpositions in ultracold bosons in a double-well potential}

\author{M.~A. Garcia-March$^{1,2,3,4}$ and Lincoln D. Carr$^{1,5}$}
\address{$^1$Department of Physics, Colorado School of Mines, Golden, CO, 80401, U.S.A. }
\address{$^2$Department of Physics, University College Cork, Cork, Ireland}
\address{$^3$Departament d'Estructura i Constituents de la Materia, Universitat de Barcelona, Barcelona, Spain}
\address{$^4$ICFO Institut de Ci\`encies Fot\`oniques, Av. C.F. Gauss, 3, E-08860 Castelldefels, Spain}
\address{$^5$Universit\"at Heidelberg, Physikalisches Institut, 69120 Heidelberg, Germany}

\begin{abstract}
We study macroscopic superpositions in the orbital rather than  the spatial degrees of freedom, in a three-dimensional double-well system. We show that the  ensuing dynamics of $N$ interacting excited ultracold bosons, which in general requires  at least eight single-particle modes and ${N+7 \choose N}$ Fock vectors,  is described by a surprisingly small set of many-body states.   An initial state with half the atoms in each  well, and purposely excited in one of them, gives rise to the tunneling of axisymmetric and transverse vortex structures.  We show that transverse vortices tunnel orders of magnitude faster than axisymmetric ones and are therefore more experimentally accessible.  The tunneling process generates macroscopic
superpositions  only distinguishable by their orbital properties and within experimentally realistic times.
\end{abstract}


\maketitle

\section{Introduction}

Orbital physics plays a crucial role in many important phenomena, like high-temperature superconductivity or colossal magnetorresistance~\cite{1998ImadaRMP}, due to the combination of orbital degeneracy and anisotropy of the vibrational states, and its correlation to other attributes, like charge or spin. 
Outstanding examples of many-body physics that can be achieved with orbital or $p$-band physics include the XYZ model and its accompanying plethora of quantum phases and transitions~\cite{2013PinheiroPRL}.
Recent experiments explored this role in the physics of ultracold atoms in three dimensional (3D) optical lattices~\cite{2011WirthNatPhys}, where this degree of freedom can be separated from those of charge and spin, and  is the origin of properties such as novel phases or supersolidity~\cite{2011LewensteinNatPhys}. Ultracold atoms are a natural system for realizing macroscopic superposition (MS) states~\cite{footnote1},  but such states have not been experimentally demonstrated, in part due to their very short decoherence times.  In this Article we propose a new kind of
MS state based on orbital properties, a \emph{vortex macroscopic superposition} (VMS), which has the potential for greatly increased decoherence times.

\begin{figure}
\includegraphics[trim =0mm 1mm 1mm 1mm, clip, width=\columnwidth]{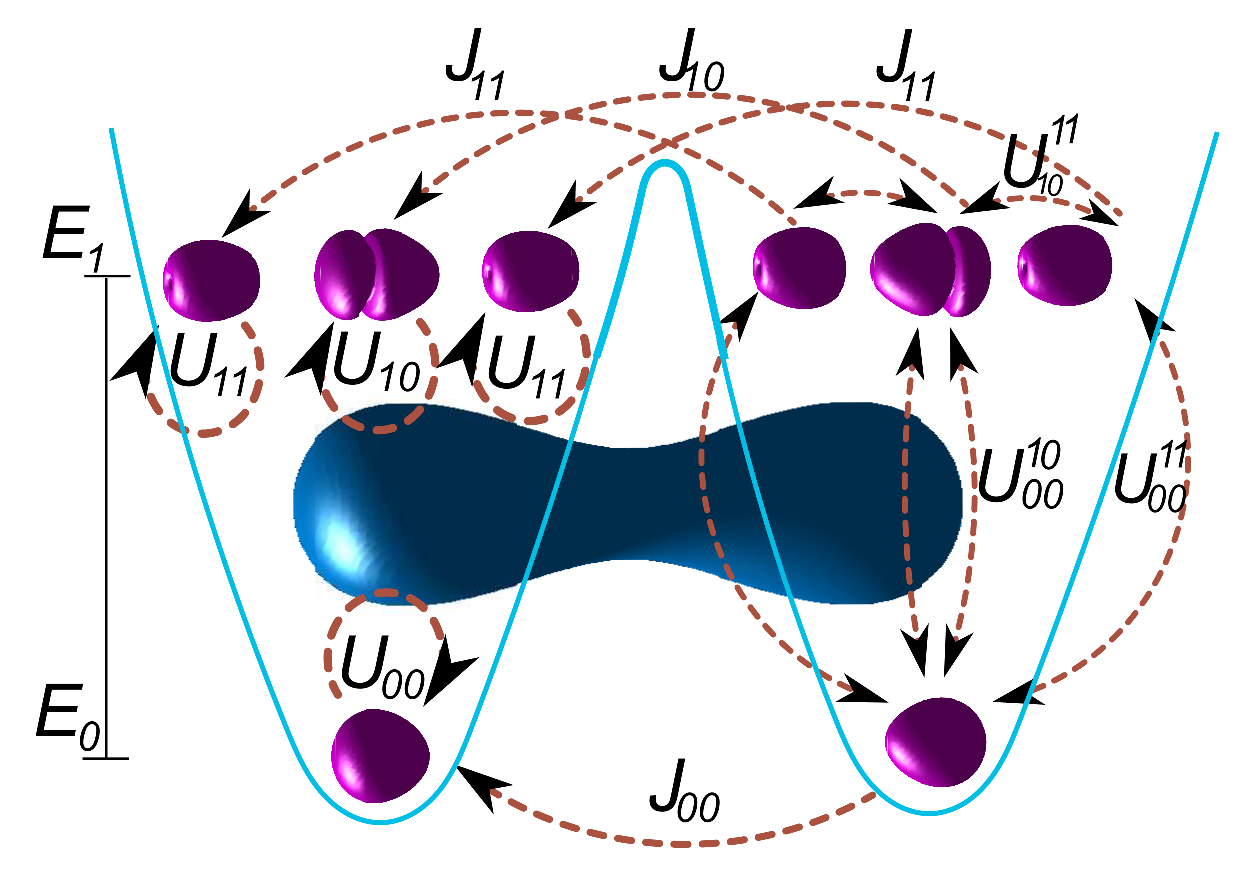}
\vspace{-0.0cm}
\caption{(Color online)  {\it Schematic  of the 3D Double Well (DW)}. The eight  modes are represented as distorted spherical harmonics (magenta) and the processes among them by arrows. In the right well, the arrows represent both tunneling and interacting processes. The blue surface is the DW equipotential surface.  \label{fig1}}
\vspace{-0.cm}
\end{figure}

Ultracold bosons in  double wells (DWs) are conventionally described by a two-mode approach, i.e., two ground modes $\ell=0$ localized in either one of the two wells. If a great part of the population is intentionally excited to the first energy level~\cite{2011WirthNatPhys}, the three degenerate $\ell=1$ orbital modes localized at each well with  $z$-component of the angular momentum $m=0,\pm1$ have to be considered, together with new processes (see Fig.~\ref{fig1}).   These excited orbital modes are vortex structures. We assume that initially  the atoms are distributed equally between both wells, and excited to an orbital mode in one of them.
Then,  the vortex tunnels between wells  with a period shorter than the lifetime of a conventional experiment, while the number occupation of both wells remains constant. This process is accompanied  by the creation of  VMSs.  Conventional spatial MSs decohere after a single interaction with an external particle.  We argue that, in contrast, VMSs must interact with many particles in order to spatially resolve the presence of the vortex in one well;  thus they are expected to be stronger against decoherence than other MSs in DWs~\cite{2010PichlerPRA}.

Ultracold bosons condensed in the ground modes in DWs  undergo two major processes: in-well interactions in pairs with energy $U$ or tunneling between wells with energy $J$. The phenomena predicted by mean-field approaches, e.g., macroscopic quantum tunneling and self-trapping~\cite{Milburn1997PRA}, were observed already in experiments~\cite{2005AlbiezPRL}. Also, it has been shown that vortices can tunnel in DWs and that vortex-antivortex MSs can be obtained in a single trap~\cite{2009SalgueiroPRA}. 
In DWs these MSs are expected in the  \textit{Fock regime} $NU\gg U \gg J$, with $N$ the number of atoms. In  this regime, mean-field approaches cease to be useful, and other methods are required,  like multiconfigurational Hartree methods~\cite{2008AlonPRA}, which are impractical in 3D DWs for VMSs.  A method based on direct diagonalization of Lipkin-Meshkov-Glick-like Hamiltonians with more than two modes is used here, because it permits  one to treat the 3D DW, to calculate with more atoms, and to obtain analytical results with perturbation techniques~\cite{2010CarrEPL}.

This Article is organized as follows. In Sec.~\ref{sec:Ham}  we introduce the eight-mode Hamiltonian and discuss the relevant processes concerning orbital modes. In Sec.~\ref{sec:tun} we discuss the dynamics  of vortex macroscopic superpositions. In sec~\ref{sec:conc} we offer our conclusions. 

 \section{Hamiltonian} 
 \label{sec:Ham}
 
 A dilute gas of $N$ ultracold bosons of mass $M$ interacting through two-body interactions with coupling constant $g=4\pi\hbar^2 a_s/M$, with $a_s$ the $s$-wave scattering length, is trapped in a 3D DW potential $V (\mathbf{r})$, consisting in a double well in the $z$ direction and harmonic transverse potentials. The latter can take any applicable functional form, for example a Duffing potential,  $V(z)=V_0(-8 z^2/a^2+16 z^4/a^4+1)$ (see equipotential surface in Fig.~\ref{fig1}). 
 We consider a two-level  eight single-particle mode expansion of the field operator in second quantization (for further details, see Ref.~\cite{2011GarciaMarchPRA}). These modes are functions localized in each well which we construct from appropriate combinations of the eigenfunctions of the single particle Hamiltonian
$H_{\mathrm{sp}}=-\frac{\hbar^{2}}{2M}\nabla^{2}+V(\mathbf{r})$. The  modes located at each well can be  described by the same quantum numbers as the spherical harmonics of a single well potential: the angular momentum
$\ell$, its $z$-component $m$, and the level index $n$. Note that, in contrast to spherical harmonics, the actual modes can be distorted in the $z$-direction (see Fig.~\ref{fig1}).
For each well, we consider only a ground mode
and the three modes at the first excited level of energies, which we name \emph{orbital modes}.
Therefore, the level index $n$
is redundant with $\ell$, and we omit it in the following. Then, the field operator can be expressed
as
\begin{equation}
\hat{\Psi}(\xvec)=\sum_{j,\ell,m}\bhat{j}{\ell}{m}{}\psi_{\ell
m}(\xvec-\xvec_{j}).\label{eq:hatPsi-1}
\end{equation}
where $\psi_{\ell m}(\xvec-\xvec_{j})$ are the modes localized
at the well denoted with index $j\in\{1,\;2\}$, whose minima is located
at position $\xvec_{j}$. Here, $\ell=0$ and $m=0$ for the ground
 mode and $\ell=1$ and $m=-1,\;0,\;1$ for the orbital modes. The operators
$\bhat{j}{\ell}{m}{}$ obey bosonic commutation relations. 
 This procedure yields the Hamiltonian~\cite{2011GarciaMarchPRA,2012GarciaMarchFiP}
 \begin{equation}
 \label{eq:allH}
\hat{H} \equiv \sum_{\ell m} \hat{H}_{\ell m} + \hat{H}_{\mathrm{int}}\,,  
 \end{equation}
with
 \begin{equation}
\label{Eq:H_bh}
\begin{array}{ll}
 H_{\ell m} &= U_{\ell m}\sum_j  \nhat{\ell}{m}{j}(\nhat{\ell}{m}{j}-1)\\
& -J_{\ell m}\sum_{j'\neq j} \bhat{j}{\ell}{m}{\dagger}\bhat{j'}{\ell}{m}{}+\sum_jE_{\ell}\nhat{\ell}{m}{j}\,,
 \end{array}
 \end{equation}
where $\hat{n}_{j\ell m}=\hat{b}^{\dagger}_{j \ell m} \hat{b}_{j \ell m}$ is the number operator.
On the other hand $\hat{H}_{\mathrm{int}} =\hat{H}_{\mathrm{inter}}^m+\hat{H}_{\mathrm{intra}}$  is given by
\begin{equation}
\label{Eq:inter}
 \begin{array}{ll}
\hat{H}_{\mathrm{inter}}^m &\equiv U_{0 0}^{1 m}    \sum_{j}  [(\bhat{j}{0}{0}{\dagger}\!)^2\,
     \bhat{j}{1}{m}{}\bhat{j}{1}{-m}{}+\mathrm{\hc}]\vspace{0cm}\\
&+ 4\,U_{0 0}^{1 m}\sum_j \,\nhat{0}{0}{j}\,\nhat{1}{m}{j},\vspace{-0cm}
\end{array}
 \end{equation}
\begin{equation}
\label{Eq:intra}
 \begin{array}{ll}
&\hat{H}_{\mathrm{intra}}\equiv  U_{1 0}^{1 1} \sum_j [(\bhat{j}{1}{0}{\dagger}\!)^2
  \bhat{j}{1}{1}{}\bhat{j}{1}{-1}{}
  +\hc ]\vspace{0.cm}\\
&+ 2\,U_{1 0}^{1 1}\sum_j\,\nhat{1}{0}{j}\,(\nhat{1}{1}{j}+\nhat{1}{-1}{j})
+2\,U_{1 1}^{1 1}\,\sum_j(\nhat{1}{1}{j}\nhat{1}{-1}{j}),\vspace{-0.cm}
 \end{array}
\vspace{.0cm}
 \end{equation}
All the coefficients in this Hamiltonian are
obtained from integrals over the on-site localized modes $\psi_{\ell
m}(\xvec-\xvec_{j})$, as follows:
\begin{equation}
J_{\ell m}\!=\!-\!\!\!\int\!\! d^{3}\xvec\,\psi_{\ell
m}^{\ast}(\xvec\!-\!\xvec_{j})\, H_{\mathrm{sp}}\,\psi_{\ell
m}(\xvec\!+\!\xvec_{j}),\label{eq:Coef_J}
\end{equation}
and
\begin{equation}
U_{\ell'm'}^{\ell m}=\frac{g}{2}\int\! d^{3}\xvec\,|\psi_{\ell
m}(\xvec)|^{2}|\psi_{\ell'm'}(\xvec)|^{2}.\label{eq:Corf_U}
\end{equation}
Note  that these coefficients are not independent. Indeed, if the modes  are
approximated by the spherical harmonics, expressions for the relationships among
them can be obtained, but the validity of these expressions is restricted to certain regimes (see
\cite{2011GarciaMarchPRA}). In obtaining  Hamiltonian~(\ref{eq:allH}) we have
 neglected the terms that involved interactions between atoms
in different wells, which are negligible for larger barrier heights, as interest
us here for the Fock regime and creation of macroscopic vortex superposition
states. We also assumed sufficiently small interactions that only single-particle modes of angular momentum up to $\ell=1$ are required.  To slightly simplify the notation we denote the interaction  coefficients for atoms with the same $\ell$ and $m$, $U_{\ell m}^{\ell m}$, as
$U_{\ell m}$. Finally,  $E_{\ell} $ is the energy  at level $\ell$. 

Hamiltonian~(\ref{eq:allH}) has a part, denoted as $H_{\ell m}$ and given by Eq.~(\ref{Eq:H_bh}), which involves only interacting and tunneling terms between atoms with the same $\ell$ and $m$. These tunneling and interaction processes are  analogous to those found in a Bose-Hubbard Hamiltonian. In contrast, the second part, given by $\hat{H}_{\mathrm{int}}$ and Eqs.~(\ref{Eq:inter}) and~(\ref{Eq:intra}), involves transitions between modes with different $\ell$ and $m$, and are the double-well analog of a Bose-Hubbard model extended to the $p$-band of a 3D optical lattice.  They are also have some terms loosely analogous to spin-1 bosons in the $s$-band Bose-Hubbard model.
Equation~(\ref{Eq:inter}) with $m=0$ accounts for   {\it zero-vorticity interlevel transitions}  which excite two atoms in the ground mode  to an orbital mode with $m=0$, or   vice versa, with an energy $U_{0 0}^{1 0}$. For  $m=\pm1$,  Eq.~(\ref{Eq:inter}) accounts for   {\it vortex-antivortex interlevel transitions}, in which two atoms are  excited to (or decay from) an orbital mode with $m\ne 0$, each with a different sign of $m$, with energy $U_{0 0}^{1 1}$.  Equation~(\ref{Eq:intra}) accounts for a third process, the  {\it vortex-antivortex intralevel transition}, in which
two  excited atoms with $m=0$ can generate a pair of atoms with $m=\pm1$ (each with different sign of $m$), or   vice versa, with energy  $U_{1 0}^{1 1}$. All tunneling, interaction and transition processes are sketched in Fig.~\ref{fig1}. Notice that the 3D DW  shows cylindrical symmetry with respect to the $z$  axis, and 
two-fold $\mathbb{Z}_2$ symmetry, or even parity, in the transverse directions.   Thus the $z$ component of the angular momentum, $\mathbb{L}_z=\sum m \hat{n}_{j 1 m}$, is conserved, while, in general, the angular momentum  $\mathbb{L}^2$ is not (see Appendix~\ref{sec:L} for an expression of this operator in second quantized language).

\section{Tunneling of vortices in a three-dimensional double well with zero population imbalance}
\label{sec:tun}

We focus   on the Fock regime, where the interactions dominate the tunneling, $\zeta_{\ell m}\equiv J_{\ell m}/U_{\ell m}\ll 1$. We further assume that $\chi\equiv N \max_{\,\ell m \ell' m'}[U_{\ell m}^{\ell' m'}]/\Delta E \ll 1$, with  $\Delta E=E_1-E_0$, so the eight-mode single-particle approximation is accurate~\cite{2011GarciaMarchPRA}. A value of $\chi$ smaller but of the order of 1 is compatible with this model if the difference with the next energy level, that is $\Delta E'=E_2-E_1$, is much larger than  $\Delta E$. This can be realized with   potential  wells which are clearly anharmonic. In the numerical examples provided we only specify $U_{00}$, $\zeta_{0 0}$, $J_{10}/J_{00}$, and $\chi$. The undetermined coefficients  $U_{\ell'm'}^{\ell m}$ are calculated by using the spherical harmonic approximation for the modes in Eq.~(\ref{eq:Corf_U}), because a relationship  between  all these  coefficients and $U_{00}$ can be obtained,  as detailed  in~\cite{2011GarciaMarchPRA}. 
We expand our states $\psi$ in terms of the Fock basis, $|\psi\rangle=\sum_{i=0}^{\Omega}c_i|i\rangle $, where the Fock space has dimension  $\Omega={N+7 \choose N}$ and $|i\rangle=\otimes_{j \ell m}|n_{j \ell m}^{(i)}\rangle$, with   $|n_{j \ell m}^{(i)}\rangle=(n_{j \ell m}^{(i)}!)^{-1/2}(\hat{b}^{\dagger}_{j \ell m})^{n_{j \ell m}^{(i)}}|0\rangle$ (see Appendix~\ref{Sec:regimes}). 

No tunneling process occurs   when all atoms are in the ground modes with zero population imbalance, i.e., the same occupation of both wells. 
If the atoms in one well are experimentally~\cite{2000MadisonPRL,2011BuckerNatPhys} orbitally excited, vortex structures tunnel between wells, creating VMSs.
 There are two cases: either (i) the vortex is {\it axisymmetric} and lies in the $z$-direction, i.e., $m=\pm 1$; or (ii) the vortex is {\it transverse}, and lies in the transverse plane showing $m=0$.  Other limiting cases are discussed in~\cite{2012GarciaMarch}.  Figures~\ref{fig:Supl_Fig1} (a) and (b) of Appendix~\ref{Sec:regimes} show  a schematic of the initial axisymmetric and transverse vortex states.

\begin{figure}
\includegraphics[width=0.95\columnwidth]{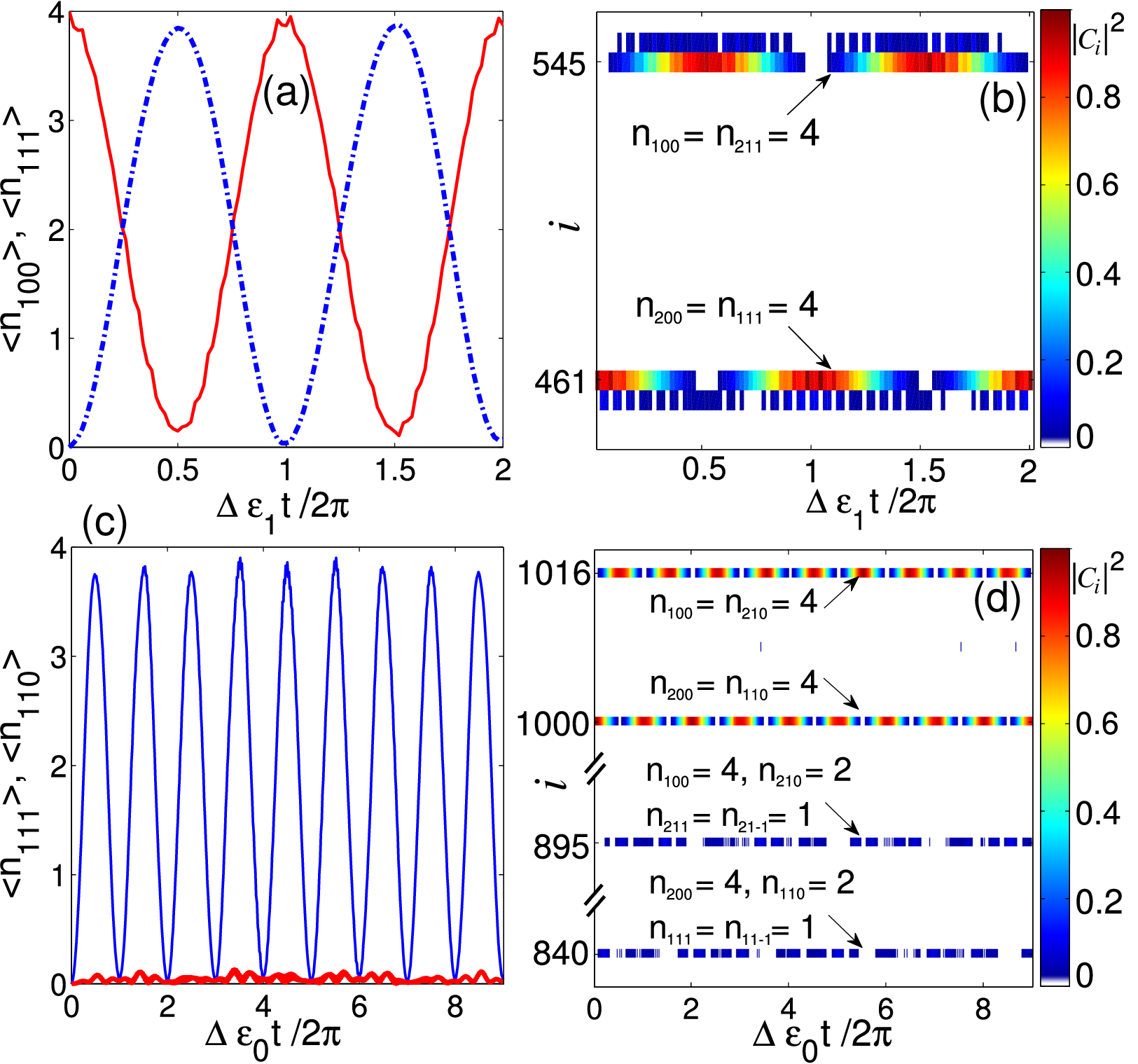}
\vspace{-.0cm}
\caption{(Color online) {\it Axisymmetric and transverse vortex tunneling}. (a) Average occupation of  well $j=1$ for unexcited atoms (red solid curve) and excited atoms in an axisymmetric vortex with $m=1$ (blue dash-dotted curve). (b) Probability densities $P_i(t)$  showing the dominant Fock vectors $i$ contributing to dynamics.  The excited and unexcited atoms slosh between wells with the same period.  (c) Average occupation of  well $j=1$ for excited atoms with $m=0$ (blue thin curve),  and  with $m=1$ (red thick curve) for the transverse vortex. (d) Probability densities  showing that two many-body Fock vectors dominate the dynamics. The small coupling to an extra pair of vectors is due to the vortex-antivortex intralevel transitions. 
\label{fig2}
}
\vspace{-0.cm}
\end{figure}

\subsection{Axisymmetric Vortex} 

The initial state is $|i\rangle $ with $n_{1 0 0}^{ (i)}=N/2$ and
$n_{2 1 1}^{(i)}=N/2$. Thus half of the atoms are localized  in well $j=1$ with $\ell=0$, $m=0$, and the other half excited in well $j=2$ to an orbital with $\ell=1$, $m=1$ [see Fig.~\ref{fig:Supl_Fig1} (a)]. We use high order perturbation theory to analyze the spectra of the two-level Hamiltonian, the perturbing part being  the hopping and transition processes. We obtain  quasidegenerate paired eigenvectors, $\psi_\pm=(1/\sqrt{2})(|i\rangle\pm|j\rangle)$, with $n_{2 0 0}^{(j)}=N/2$  and $n_{1 1 1}^{(j)}=N/2$,   and splitting given by
\begin{align}
\label{Eq:splittingN/2}
&\Delta\varepsilon_{\pm1}=2(N/2!)^2 J_{00}^{N/2}J_{11}^{N/2}\tilde{U}_a\,, \\
 &\tilde{U}_a=\sum_{i_{1}=0}^{1}\sum_{i_{2}=0}^{1}\!\dots\!\!\!\sum_{i_{N-1}=0}^{1} \prod_{j=1}^{N-1}U_a\left(\sum_{k=1}^{j}i_{k},j-\sum_{k=1}^{j}i_{k}\right)\,,\nonumber\\
&U_a(n,p)=(-1)^{(n+p)}\left\{U_{00} 2n[N/2-n]+U_{11} 2p[N/2-p]\right.\nonumber\\
&- \left. 4U_{00}^{11}\left[N/2(n+p)-2np\right]\right\}^{-1}\,,\nonumber
\end{align}
with  $N/2-1\le\sum_{j=1}^{N-1}i_{j}\le N/2$ (see Appendix~\ref{sec:pert} for the derivation of this expression). We validated Eq.~(\ref{Eq:splittingN/2}) for small $N$ in simulations, and note that, despite the complex sums and lack of a simple expansion, the essential scaling is $\Delta\varepsilon_{\pm1} \sim U_{00}(J_{00}J_{11})^{N/2}/[N!(U_{00})^N]$, since $U_{00}\sim U_{00}^{11} \sim U_{11}$ up to factors of order unity.

\begin{figure}
\includegraphics[width=0.95\columnwidth]{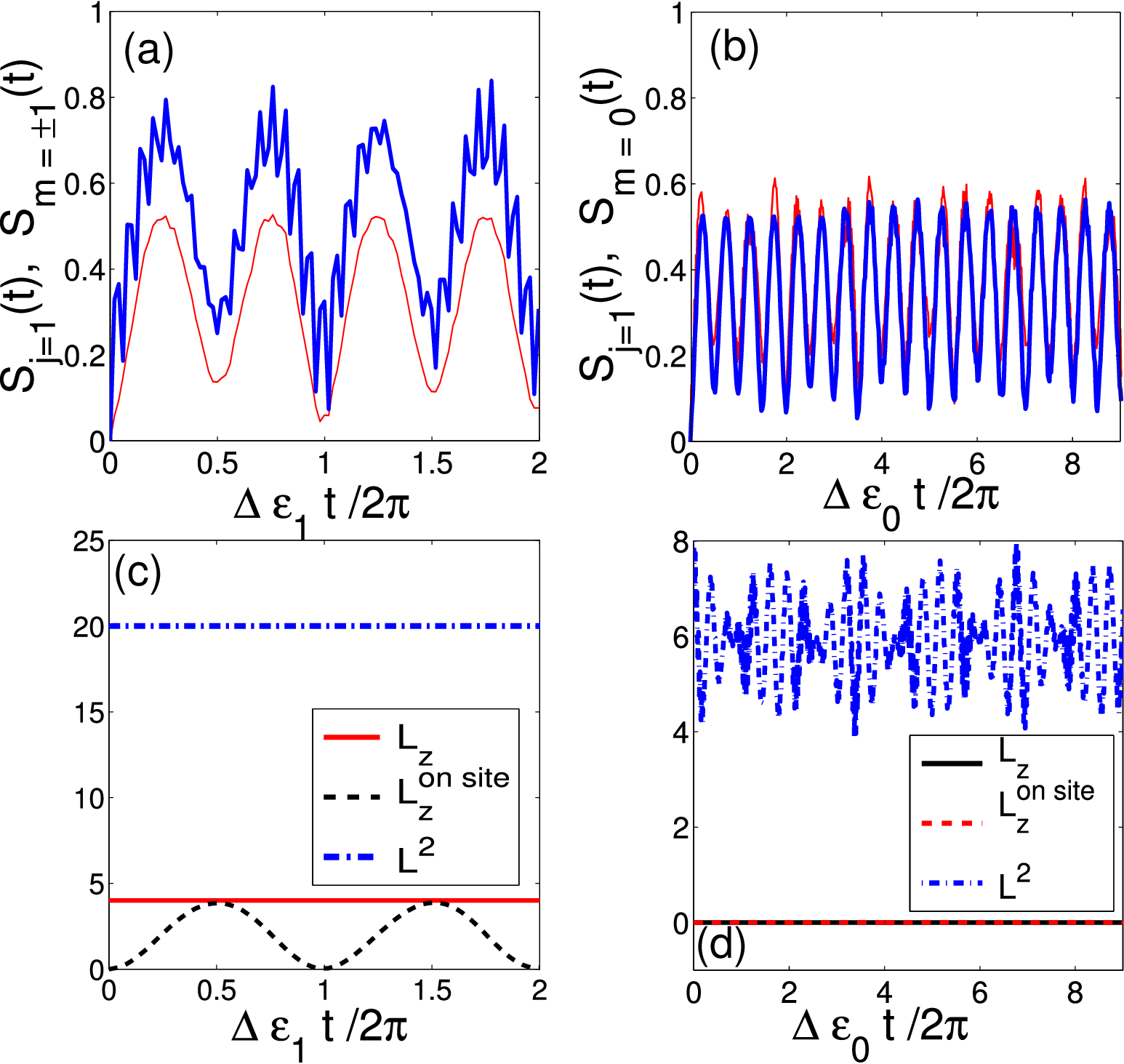}
\vspace{-0.cm}
\caption{(Color online) {\it  Entropies and angular momentum for the axisymmetric and transverse vortices}.
(a) Spatial (blue thick curve) and angular momentum  for $m=1$ (red thin curve)  entanglement entropy normalized to their maximal possible value, for the axisymetric vortex.  (b) Same for the transverse vortex, with  angular momentum entropy for $m=0$.   (c)  For the axisymmetric vortex, the total  $\mathbb{L}^2$ (blue dash-dotted curve) and $\mathbb{L}_z$ (red solid curve)  are conserved, while on-site $\mathbb{L}_z$ (black dashed curve) is not. (d) For the transverse vortex, both total $\mathbb{L}_z$ and on-site $\mathbb{L}_z$ (superposed) are conserved, while total $\mathbb{L}^2$ is not.  \vspace{-0.cm}\label{fig3}}
\end{figure}

The system oscillates between states $|i\rangle $  and $|j\rangle$, with period $ T=2\pi\hbar/\Delta \varepsilon_{\pm1}$. Thus, half of the atoms remain non-excited and the other half excited with $m=1$, both populations sloshing between wells with periodic average occupations,   $\bar{n}_{j \ell m}(t)=\left\langle \psi(t)|\hat{n}_{j \ell m}|\psi(t)\right\rangle $.  The quantum dynamics calculated via numerical exact diagonalization~\cite{footnote2} is shown in Fig.~\ref{fig2}(a),  for $U_{00}=1$, $\zeta_{0 0}=1/5$, $J_{10}/J_{00}=5$, and $\chi=1/10$.  We simulated from $N=1$ to 12, and we discuss in Appendix~\ref{Sec:regimes} the time cost of our algorithm as $N$ is increased. Here we illustrate $N=8$ as the case of unit filling (one atom per mode) is intriguing.  We emphasize that our analytical results are valid for arbitrary $N$. The probabilities  $ P_{2 1 1}(t)$ of finding $N/2$ excited   atoms in well $j=2$ with $m=1$ and   $P_{1 0 0}(t)$ of finding the other 
$N/2$ non-excited in well $j=1$ are equal,  $P_{2 1 1}(t)=P_{1 0 0}(t)=\cos^2(\Delta \varepsilon_1 t/2\hbar)$.  At the quarter period $P_{2 1 1}(t)=P_{1 1 1} (t)=1/2$ and similarly for the atoms in the ground modes. Then, the half of  initially  excited atoms  with $m=1$ and the non-excited other half have, at $T/4 $,  equal probability to occupy both wells.  This is a VMS in the $z$-component of the angular momentum. Figure~\ref{fig2}(b) plots the probability density in time, $|c_i(t)|^2$, for the Fock vectors labeled with index $i$. Here all
$|c_i(t)|^2$ are negligible in time, except for $i=461$ and  $j=545$, corresponding to the vectors $|i\rangle $ and $|j\rangle $, respectively (see  Appendix~\ref{Sec:regimes} for the ordering of index $i$). At $t=T/4$ both coefficients  are $1/\sqrt{2}$, showing that the system is in a VMS. In the numerical results depicted in Fig.~\ref{fig2}(b) there is a small coupling to two nearby vectors in Fock space,  one with  $n_{1 0 0}^{(i')}=n_{2 1 1}^{(i')}=N/2-1$ and $n_{2 0 0}^{(i')}=n_{1 1 1}^{(i')}=1$ while the other has $n_{2 0 0}^{(j')}=n_{1 1 1}^{(j')}=N/2-1$ and $n_{1 0 0}^{(j')}=n_{2 1 1}^{(j')}=1$. The reason is that, for the parameters chosen, the tunneling energy is not much smaller than the energy gap with the quasi-degenerate pair $\psi'_\pm=(1/\sqrt{2})(|i'\rangle\pm|j'\rangle)$, which is of the order of $U_{00}$.  Then, this coupling is highly suppressed as  $\zeta_{0 0}$ is reduced.  Since we consider
only pure states, the VMS has zero total quantum von Neumann entropy.  However, the local entanglement or von Neumann entropy both in space and angular momentum are non-zero: the partial trace over the four modes in  well $j=1$, or alternately, over all modes but $\ell=1,m=\pm 1$, yields algebraically complex expressions for $S_{j=1}$ and $S_{m=\pm 1}$, not shown here for brevity~\cite{2012GarciaMarch}.  We normalize both entropies to their maximal possible value, which coincide because we trace over the same number of modes,  and plot the results in Fig.~\ref{fig3}(a).  Both show a maximum at $ T/4$, when the VMS occurs. Finally, both   $\mathbb{L}^2$ and   $\mathbb{L}_z$ are conserved, while the on-site $\mathbb{L}_z$, where $j$ is restricted to one well, oscillates with period $T$ [see Fig.~\ref{fig3}(c)].

\begin{figure}
\includegraphics[width=0.65\columnwidth]{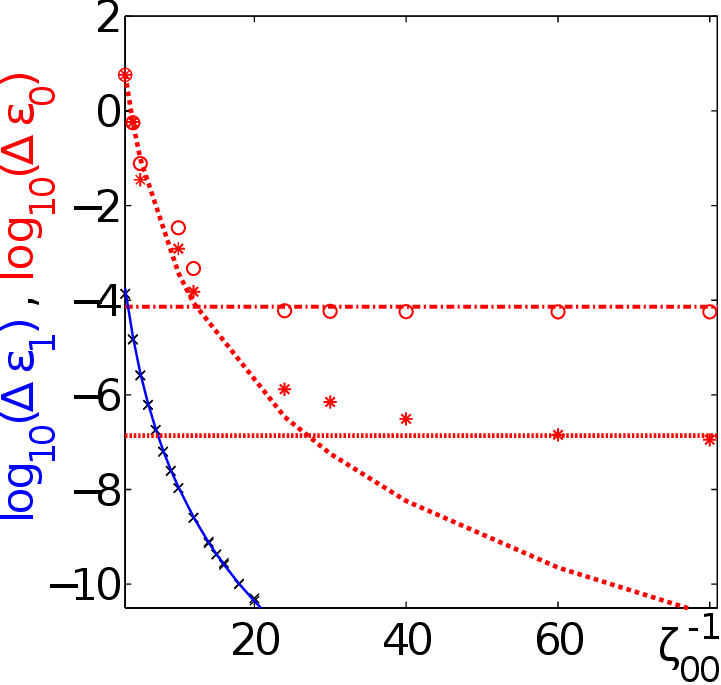}\vspace{-0.cm}
\caption{(Color online) {\it Energy splittings for vortex macroscopic superpositions (VMSs).}   Axisymmetric (analytical: solid blue curve; numerical: blue crosses) and transverse [analytical: dashed (small $1/\zeta$ limit), dotted (large $1/\zeta$ limit with $\chi=1/80$), and dash-dotted (same with $\chi=1/10$) red curves; numerical:  asterisks ($\chi=1/80$) and circles ($\chi=1/10$)] energy splittings.  In the transverse case there are two  limits: small and  large $1/\zeta$. The analytical small $1/\zeta$ limit does not depend on $\chi$.  The analytical large  $1/\zeta$ limit depends on $\chi$.
\vspace{-0.cm} \label{fig4}}
\end{figure}

\subsection{Transverse Vortex} 

  Initially the excited atoms in one well  have $m=0$, i.e., the initial state $|i\rangle $ has $n_{2 0 0}^{(i)} =n_{1 1 0}^{(i)}=N/2$ [see Fig.~\ref{fig:Supl_Fig1} (b)]. Then, the intralevel vortex-antivortex transitions  create atoms   with $m=\pm 1$ from excited atoms with $m=0$. Using perturbation theory we find that the relevant Fock vectors include not only $|i\rangle$ and $|j\rangle$ (with $n_{1 0 0}^{(j)} =n_{2 1 0}^{(j)} =N/2$), but also, due to this   process, the vectors $|k\rangle$ and $|l \rangle$, with  $n_{1 0 0}^{(k)} =N/2$, $n_{2 1 0}^{(k)} =N/2-2$,  and  $n_{2 1 \pm 1}^{(k)} =1$; and with $n_{2 0 0}^{(l)} =N/2$, $n_{1 1 0}^{(l)} =N/2-2$,  and  $n_{1 1 \pm 1}^{(l)} =1$ [see Appendix~\ref{sec:pert} and Fig.~\ref{fig:Supl_Fig1} (c)].  Thus despite ${N+7 \choose N}$ Fock vectors, the dynamics is dominated by combinations of just four of them, the quasidegenerate pairs $\psi_{\pm} =\alpha|i\rangle\pm \alpha|j\rangle+\beta|k\rangle\pm \beta|l\rangle$, with splitting  $\Delta\varepsilon_0 $,
and $\phi_{\pm} =\mp \beta|i\rangle - \beta|j\rangle\pm \alpha|k\rangle+ \alpha|l\rangle$,  with splitting $\Delta\varepsilon_0'$.  Perturbation theory shows that $\alpha\gg\beta$ and all couplings to other Fock vectors are negligible~\cite{2012GarciaMarch}.
We find the average number of atoms in the  well $j=2$ in a transverse vortex  to be
%
$\left\langle n_{200}\right\rangle = 2C(1-\cos\Delta \varepsilon_0 t)+2C'(1-\cos\Delta \varepsilon'_{0} t )\,,$
where $\hbar=1$, $C=\alpha^{2}N(\alpha^{2}+\beta^{2})$, $C'=\beta^{2}N(\alpha^{2}+\beta^{2})$,
and   $C\gg C' $ since $\alpha\gg\beta$. In Fig.~\ref{fig2}  we present the evolution of (c) the average occupation of well $j=1$ for excited atoms either with $m=0$ or $m=1$,  and (d)  the probability density $|c_i(t)|^2$, when $U_{00}=1$, $\zeta_{00}=10^{-2}$, $J_{10}/J_{00}=5$, and $\chi=1/10$. Even deep in the Fock regime (very small  $\zeta$), $\beta$ remains around  5$\%$ of $\alpha$.  This coupling to the pair  $\phi_{\pm}$ is due to  the vortex-antivortex intralevel transitions. As this process depends on the interactions, it is not negligible even for very small $\zeta$,  in contrast to Fig.~\ref{fig2} (b).  
This four eigenvector problem leads to a quasiperiodic motion in which the two relevant  frequencies are proportional to the splittings, with a very small modulation due to the small  coupling to   $|k\rangle$ and $|l \rangle$.  The von Neumann entropies $S_{j=1}$ and  $S_{m = 0}$,  the latter obtained from the partial trace over all modes but $\ell=1,m=0$,  are shown in Fig.~\ref{fig3}(b). Both are normalized to their maximal possible value, which for $S_{m = 0}$ differs from $S_{j=1}$ and $S_{m=\pm 1}$, because we trace only over two modes.  Now, the atoms tunnel  between both wells with the same period, creating a
transverse VMS, and  vortex-antivortex pairs are rapidly created and annihilated during evolution.  Finally,  while $\mathbb{L}^2$ is not constant,  both total and on-site $\mathbb{L}_
z$ are conserved, since the atoms with $m=\pm1$ are  created in pairs [see Fig.~\ref{fig3}(d)].

The splittings can be obtained using perturbation theory. There are two possible scenarios.  First, if $\zeta_{00}$ is bigger than $\chi$,  the splitting is given by Eq.~(\ref{Eq:splittingN/2}) upon substitution of  $J_{11}$ by  $J_{10}$, $U_{11}$ by $U_{10}$, and $U_{00}^{11}$ by $U_{00}^{10}$. Second, if  $\zeta_{00}$ is much smaller than $\chi$, the coupling between $|i\rangle$ and $ |j\rangle$ is due to the zero-vorticity interlevel transitions. Then, this splitting is
\begin{align}
\label{eq:splittingN2t}
& \Delta\varepsilon_0=2(N/2)\left(N/2!\right)^2 \left(U_{00}^{10}\right)^{N/2}\tilde{U}_t\,, \\
&U_t(n,p)=\{U_{00} [4n(N/2-n)-2n-2p(2p-1)]+\nonumber\\
&  U_{10}[4p(N/2-p-\frac{1}{2})-2n(2n-1)]
 +2(p-n)\Delta E\}^{-1}\,,\nonumber
\end{align}
with $\tilde{U}_t=\tilde{U}_a$ from Eq.~(\ref{Eq:splittingN/2}) with $U_a(n,p)$ replaced by $U_t(n,p)$, and the $i_j$ running only to $j=N/2-1$. Here, the essential scaling is $\Delta\varepsilon_{0} \sim U_{00}(N/2)!(U_{00}^{11})^{N/2}/[N(N-1)(\Delta E\,U_{00})^{N/4}]$.

\subsection{Comparison between the tunneling of axisymmetric and transverse vortices}

In Fig.~\ref{fig4}  we show the splitting, both for the axisymmetric and transverse vortices, for $N=8$ and different  $\zeta_{00}$.  The axisymmetric splitting is smaller than the transverse splitting, and the analytical approach shows good agreement with the numerical calculation: thus transverse vortices tunnel faster; moreover, an axisymmetric vortex will require stirring or phase imprinting the bosons localized in one well~\cite{2000MadisonPRL}, which may encounter practical difficulties, as it has to be stirred in the axial direction of the 3D DW double-well to generate the vortex  only in one well. On the contrary, the transverse vortex only requires vibrating  one well in the transverse direction, as is done in the experiments reported in~\cite{2011BuckerNatPhys}, which make it technically easier to make in an experiment. Note also that the potential wells realized in these experiments are anharmonic, thus permitting the eight-mode Hamiltonian to be valid for larger values of $\chi$. 
We also plot the analytical calculation for the transverse vortex splitting. 
 The numerical calculation shows good agreement with the curve obtained from Eq.~(\ref{Eq:splittingN/2}) for  small $1/\zeta_{00}$. For large $1/\zeta_{00}$, the splitting tends to an asymptotic value, given by Eq.~(\ref{eq:splittingN2t}), which is  increased for larger values of $\chi$.  Let us remark that the coupling between $|i\rangle$ and $ |j\rangle$ only due to the zero-vorticity interlevel transitions occur when $N/2$ is even. For $N/2$ or $N$ odd, the coupling requires also the
tunneling process, and since in this regime $J_{\ell m}$ is very small,  $\Delta\varepsilon_0 $ is much smaller than the one given by Eq.~(\ref{eq:splittingN2t}). For example, for $\zeta=10^{-2}$ and $\chi=1/10$, we  obtain $\Delta\varepsilon_0 = 5.7\times 10^{-5}$ for $N=8$, while $\Delta\varepsilon_0 =3.8\times 10^{-9} $ and $\Delta\varepsilon_0 =3.2\times 10^{-11} $ for  $N=7$ and  $N=9$, respectively.

\subsection{Experimental feasibility}

Let us obtain the period of oscillation for typical experiments  with $\omega=2\pi\times 70$ Hz to $ 7$ Khz  with $\Delta E= \hbar\omega$. Taking $\chi=1/2$  we obtain $U_{00}=\chi\Delta E/N =(0.125 \,\mathrm{to}\,  12.5)$ Khz, which for $\zeta_{00} =1/100$ gives $J_{00}=(1.25\,\mathrm{to}\,  125)$ Hz. Then we obtain $\Delta \varepsilon\approx7\times (10^{-3}\,\mathrm{to}\, 10^{-1})$ kHz, which gives a period of oscillation  $T=2\pi\hbar/\Delta \varepsilon=1 \,\mathrm{to}\,  0.01 $ s (an oscillation frequency of  $1 \,\mathrm{to}\,  100 $ Hz).
MSs will be observable in an experiment if this time is shorter than the decoherence time. Conventional NOON states, where all atoms occupy simultaneously both wells, are  fragile against decoherence processes, e.g. induced by imperfections of the potential~\cite{2010CarrEPL}, spontaneous emission, or the thermal cloud~\cite{2010PichlerPRA}, since they decohere after a single interaction.  A thorough study of decoherence will require the solution of a Master equation, which is out of the scope of this paper. Nevertheless, conventional environments and their interaction with the system do not include terms that distinguish between angular properties, but only densities. 
This indicates longer decoherence times, as the vortex core has to be resolved to make the VMS collapse, this core being the volume at which the vortex single-particle eigenfunction is negligible.  The number of interactions is proportional to the total core volume times the number density of the
condensate in that region.  For example, scaling up to a larger condensate, for typical condensate densities of $10^{13} \mathrm{cm}^{-3}$, and a core area of a healing length of $(0.5 \mu\mathrm{m})^2$ times a transverse dimension of 10 $\mu$m the VMS decoherence time will be 125 times larger than for conventional MSs.  For our 8 atom case we expect decoherence times to be at least 3 to 4 times longer. 
 The back-of-the-envelope calculation
provided here suggesting that VMSs have significantly longer
decoherence times than conventional NOON states requires
further study with nonequilibrium, open-quantum-system
methods to provide quantitative predictions and account for all
forms of decoherence and measurement. For example, small
differences in the relative cloud size between wells may also
lead to some level of distinguishability, which nevertheless we
expect to be much better than for conventional NOON states.

\section{Conclusions}
\label{sec:conc}

%
In conclusion, we have shown that an initial homogeneous distribution of atoms in a double well potential, excited in one of them to an orbital with $m=\pm1$ (axisymmetric vortex) or $m=0$ (transverse vortex)  evolves in time to vortex macroscopic superposition states which are only distinguishable by their angular properties. 
The possibility of observing these superpositions in experimentally realistic conditions required that the interactions were large enough as to make the transitions governed by the interactions $U_{00}^{10}$ for the transverse vortex to be the more relevant process in the system. We noted that, in the Fock regime considered here, the tunneling terms $J_{00}$ and $J_{10}$ are too small to generate these superpositions dynamically. This required that the energy difference between levels, $\Delta E$, had to be small enough when compared with $U_{00}^{10}$. Also, to make this transition possible in realistic times, the number of atoms $N$ cannot be so large as to make the period of oscillation $T$ too long.  As an example, we detailed some possible values of all the parameters in a realistic experiment which will permit to observe this superposition.
This is a new route for the realization of macroscopic superposition states with ultracold atoms in double wells with potentially much longer decoherence times. 


\acknowledgements

We acknowledge useful discussions with D. R. Dounas-Frazer and M. K. Oberthaler.  This work was supported by the NSF and the Heidelberg Center for Quantum Dynamics. MAGM acknowledges support from Fulbright commission  (USA)  and MEC (Spain) and  ERC Advanced
Grant OSYRIS (led by M.L.), EU IP SIQS, EU STREP EQuaM, John Templeton Foundation, and Spanish Ministry Project FOQUS. MAGM also acknowledges support from project  FIS2011-24154, DGI (Spain) and Generalitat de Catalunya Grant No. 2014SGR-401.

\appendix

\section{Total and \textit{z}-component angular momentum operators}
\label{sec:L}

 The operator for the  $z$-component of the angular momentum is given by
\begin{equation}
\label{eq:Lz}
\hat{\mathcal{L}}_{z}=\sum_{j,m} m\hat{n}_{j 1 m}\,,
\end{equation}
Notice that it is not necessary to sum over $\ell$ because  $m=0$ for
$\ell=0$. Thus, $\hat{\mathcal{L}}_z$ is given by the total number of particles with
$m\ne0$ in
both wells. The projection of the angular momentum on the $z$ axis is conserved
in the system, $\left[\hat{H},\hat{\mathcal{L}}_{z}\right]=0$,
due to the cylindrical symmetry of the double well with respect to the $z$
axis. This conservation means processes in the Hamiltonian changing $m$ for single atoms
must do so in pairs.  For example, the vortex-antivortex inter- and intralevel
transitions  create/annihilate a pair of atoms with $m\ne0$, one with $m=1$ the
other with $m=-1$ [see Eqs.~(\ref{Eq:inter}) and~(\ref{Eq:intra})]. Note that, if the initial state is a Fock vector with an odd number of atoms with $m\ne0$, this will only imply an initial odd value of $\hat{\mathcal{L}}_z$ which will be conserved under time evolution.

On the other hand, the total angular momentum operator can be expressed as
\begin{equation}
\hat{\mathcal{L}}^{2}=\hat{\mathcal{L}}_{z}^{2}+\frac{1}{2}\left(\hat{\mathcal{L}}_{+}\hat{\mathcal{L}}_
{-}+\hat{\mathcal{L}}_{-}\hat{\mathcal{L}}_{+}\right)\,,
\end{equation}
where the ladder angular momentum operators $\mathcal{L}_{\pm}$ are
\begin{eqnarray}
\hat{\mathcal{L}}_{+} & = &
\sqrt{2}\sum_{j}\hat{b}_{j11}^{\dagger}\hat{b}_{j10}+\sqrt{2}\sum_{j}\hat{b}_{
j10}^{\dagger}\hat{b}_{j1-1}\\
\hat{\mathcal{L}}_{-} & = &
\sqrt{2}\sum_{j}\hat{b}_{j1-1}^{\dagger}\hat{b}_{j10}+\sqrt{2}\sum_{j}\hat{b}_{
j10}^{\dagger}\hat{b}_{j11}\,,
\end{eqnarray}
and then
\begin{align}
\label{eq:L2}
&\hat{\mathcal{L}}^{2}  =  \left(\sum_{j m=\pm1}\hat{n}_{j1m}\right)^{2}+
\sum_{j m=\pm1}\Big[\hat{n}_{j00}\left(1+\hat{n}_{j1m}\right)\nonumber\\
 &+ \hat{n}_{j1m}\left(1+\hat{n}_{j10}\right)\Big] + 2\sum_{m=\pm1} \left[
\hat{b}_{110}^{\dagger}\hat{b}_{210}^{\dagger}\hat{b}_{11m}\hat{b}_{21-m}\right.
\nonumber\\
&
+\hat{b}_{110}^{\dagger}\hat{b}_{21m}^{\dagger}\hat{b}_{11m}\hat{b}_{21m}\Big]
+2\sum_{j}\left(\hat{b}_{j10}^{\dagger}\right)^{2}\hat{b}_{j11}\hat{b}_{j1-1}
+\mathrm{h.c.}
\end{align}
Due to the symmetry of the potential $V(\mathbf{r})$, the total angular momentum is not conserved,
$\left[\hat{H},\hat{\mathcal{L}}^{2}\right]\ne0$.
The $z$ component of the angular momentum in well $j$ is
\begin{equation}
\label{eq:Lz1}
 \hat{\mathcal{L}}_{z,j}=\sum_m m\hat{b}_{j 1 m}^{\dagger}\hat{b}_{j 1 m},
\end{equation}
while the total angular momentum operator of a single well (under our approximation of only two levels) is~\cite{2010TsatsosPRA}:
\begin{align}
\label{eq:Lj2}
& \hat{\mathcal{L}}_j^{2} =
 \left(\hat{n}_{j11}-\hat{n}_{j1-1}\right)^{2}+\sum_{m=\pm1}\Big[\hat{n}_{j10}
\left(1+\hat{n}_{j1m}\right)\nonumber\\
 & +\hat{n}_{j1m}\left(1+\hat{n}_{j10}\right)\Big]+
2\left(\hat{b}_{j10}^{\dagger}\right)^{2}\hat{b}_{j11}\hat{b}_{j1-1}+\mathrm{
h.c.}\,.
\end{align}

\section{Regimes, Fock vectors, and cases of study }
\label{Sec:regimes}

In this paper we focus on the Fock regime, that is, we assume $\zeta_{\ell
m}=J_{\ell m}/U_{\ell m}\ll1$ for all $\ell$ and $m$ considered.
We also assume that the separation between the two levels is bigger
than the interaction energies of the atoms, $\chi\equiv N\max_{\,\ell
m\ell'm'}[U_{\ell m}^{\ell'm'}]/\Delta E\ll1$, in order to avoid exciting higher orbital modes. 
We consider a separable potential $V(\mathbf{x})=V(x)+V(y)+V(z)$, where the 1D
potentials in the $x$ and $y$ directions are harmonic ones, and the 1D potential in the $z$
direction is a double well. This potential is characterized by a barrier height $V_0$ and
a distance between wells $a$ (we use this notation because, very loosely speaking, the double well can be thought of as a two-site lattice, and the orbital modes as Wannier functions in the $p$-band). These two parameters, together with the coupling constant, determine all
coefficients $J_{\ell m}$, $U_{\ell'm'}^{\ell m}$, and the energy difference
between levels $\Delta E$. For our calculations we consider a Duffing potential
in the $z$ direction, $V(z)=V_{0}(-8z^{2}/a^{2}+16z^{4}/a^{4}+1)$.

\begin{figure}
\includegraphics[scale=0.35]{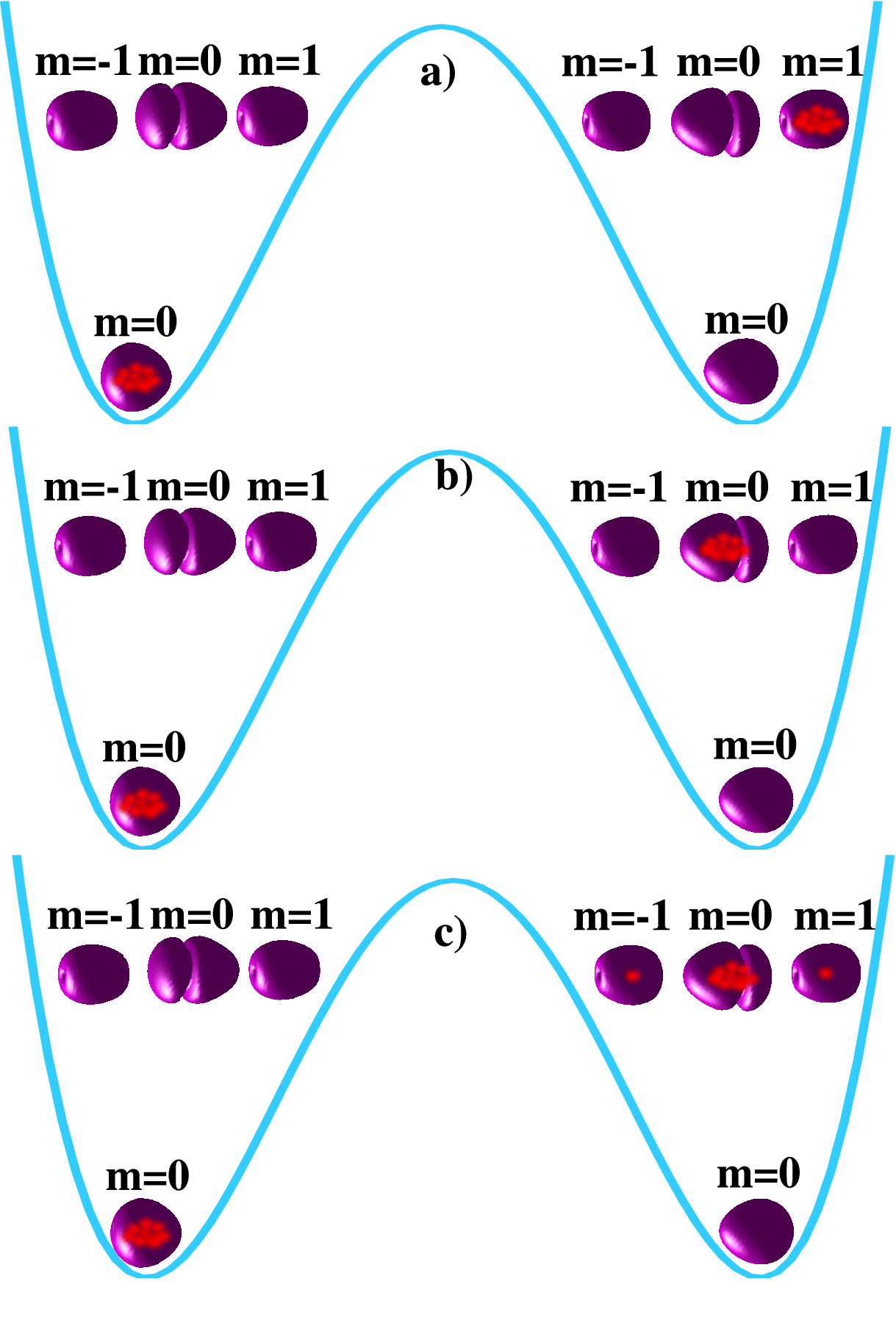}
\caption{\label{fig:Supl_Fig1} Schematic of the Fock vectors. Red circles
represent atoms. (a) and (b) represent the Fock vectors that correspond to the
initial conditions for the axisymmetric and transverse cases, respectively. (c)
represents the Fock vector
to which the transverse vortex is coupled along evolution due to the
creation/annihilation of vortex/antivortex pairs. }
\end{figure}

Let us consider the Fock vectors
\begin{equation}
\label{eq:fockv}
|i\rangle=\bigotimes_{j\ell m}|n_{j\ell
m}^{(i)}\rangle,
\end{equation}
with $|n_{j\ell m}^{(i)}\rangle\equiv(n_{j\ell m}^{(i)}!)^{-1/2}(\hat{b}_{j\ell
m}^{\dagger})^{n_{j\ell m}^{(i)}}|0\rangle$.
These Fock vectors account for all possible combinations of $N$ atoms
in the eight modes. The number of combinations is $\Omega=[(N+7)!]/[(N!)(7!)]$, and therefore
the dimension of the corresponding Hilbert space.  Considering the binomial coefficient, one observes that typical expansions in powers of $N$ using Stirling's approximation
of ${N+m \choose N}$ only become reasonably accurate for $N \gg m^2$, which is not the case in systems we consider; therefore the dimension is best expressed by the choose symbol itself. Then, we expand the ground state $|\psi\rangle$ of the
two-level eight-mode Hamiltonian in terms of these Fock vectors
$|\psi\rangle=\sum_{i=0}^{\Omega}c_{i}|i\rangle$.
We choose the Fock index $i$ to increase with the occupation of well $j=1$ of
the  ground mode, and then with the occupation of orbital modes. Then,  the
first $N+1$ Fock vectors have index $i=1+n_{100}$. Only one atom occupies the
orbital modes for the next the next $6N$ vectors, which is the total number of
combination of $N-1$ atoms in both wells in the ground mode and one atom
orbital modes with $m=-1,\,0,\,1$, in two wells. For this first set of states excited to the orbital modes, the Fock index is
\begin{align}
  i=& 2+n_{100}+N\left[\sum_m n_{11m}+(2N_{10}+4N_{11}+1)\right]\nonumber
\end{align}
where $N_{\ell m}$ is the number of atoms at level $\ell$ with z-component of
the angular momentum $m$. The Fock index $i$ increases further  with all
combinations of $N_e=2,\hdots,N$ atoms occupying the orbital modes and $N-N_e$
atoms in the ground modes.

Under these conditions we study the dynamics when initially half of
the atoms are in the ground mode located in  one of the wells, while the other half
is in the other well occupying  an orbital mode with $m=\pm1$ (axisymmetric
vortex case) or with $m=0$ (transverse vortex case). The two possible
initial states, which correspond to Fock vectors,  are schematically represented in Fig.~\ref{fig:Supl_Fig1}
(a) and (b), respectively. Our numerical results are obtained after direct
diagonalization of Hamiltonian~(\ref{eq:allH}).

\begin{figure}
\includegraphics[scale=0.55]{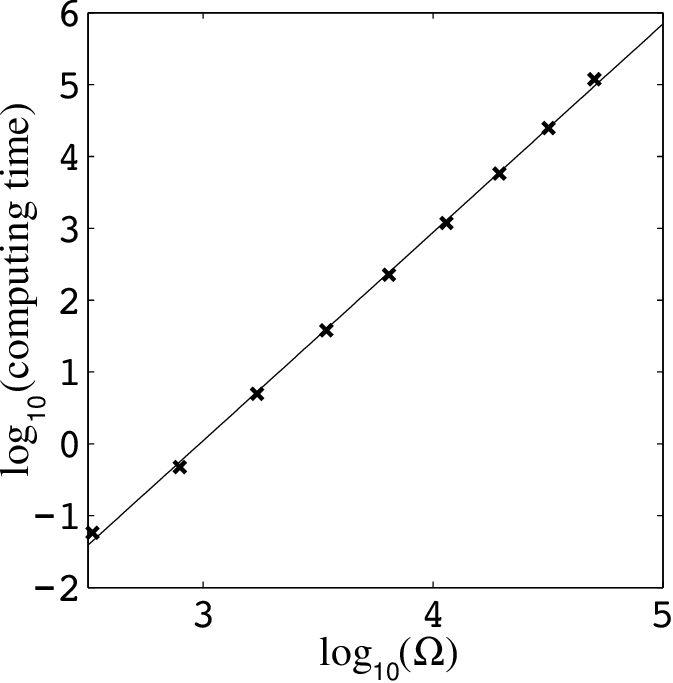}
\caption{\label{fig:Supl_Fig2} Logarithm of the time cost (in seconds) of the exact diagonalization algorithm
as a function of the logarithm dimension of the Hilbert space $\Omega$. Crosses correspond
to the numerical calculation for $N=4$ to $N=12$ atoms. Solid line corresponds
to the fitting to a straight line of slope 2.9, thus showing the
$\BigO{\Omega^3}$ behavior. }
\end{figure}

The dimension of the Hamiltonian is $\Omega^2$, and exact diagonalization has a compute time cost of $\Omega^3$.  As we prediagonalize our Hamiltonian matrix elements and then exponentiate, time evolution requires just the number of time steps $N_t$.  Then the total compute time is $\Omega^3 N_t$.  We also enforce parity of our eigenstates, as is vital in the Fock regime, where the splitting between symmetric and antisymmetric near-degenerate eigenstates is exponentially small and far beneath the computer's numerical resolution.  In Fig.~\ref{fig:Supl_Fig2} we show the actual time cost of our algorithm for $N=2$ to $N=12$ (not including the trivial time evolution cost $N_t$), and fit a polynomial in $\Omega^3$; deviations from the $\Omega^3$ scaling are due to other operations in our code, including enforcement of parity.  In the paper we focused on unit filling $N=8$ as the most interesting case (here one particle per mode, not just per site), as this is often a good starting place for lattice-type problems,
 but we have simulated all possibilities from $N=2$ to $N=12$, as we will present in future work.  We emphasize that although our Hamiltonian is sparse, complete diagonalization even of a sparse matrix does require $\Omega^3$ operations; an efficient use of Lanczos or other methods to obtain just a few eigenmodes would not obtain the highly excited states we need for our dynamics.  One could consider building on our perturbation theory to include successive sets of highly excited eigenstates of correct parity, develop a reduced effective basis, and thereby numerically reach large numbers of atoms $N \gg 12$. Such an approach may or may not be effective; we know from matrix-product-state methods that when dealing with a quantum quench, for example, perturbation theory will not suffice, and it is generally necessary in quantum entangled dynamics to have a time-adaptive approach, not a fixed basis.  Multiconfigurational Hartree methods offer another possibility, in which we could relax the requirement of only
occupying up to the $\ell=1$ single-particle first excited orbital modes; however, such methods can scale badly in 3D.  We will consider an optimized perturbative approach as well as other numerical approaches for the dynamics in future work.  For the purposes of this paper, straightforward exact diagonalization suffices to examine the unit filling case.

\section{Perturbation theory}
\label{sec:pert}

In the Fock regime $\zeta_{\ell m}=J_{\ell m}/U_{\ell m}\ll1$
we can consider all tunneling terms in Eq.~(\ref{Eq:H_bh}), that
is
\begin{equation}
\hat{H}_{J}=\! J_{\ell m}\!\!\sum_{j'\neq
j}\!\Big[\bhat{j}{\ell}{m}{\dagger}\bhat{j'}{\ell}{m}{}+\mathrm{h.c.},\label{eq:PertH_J}\Big]
\end{equation}
as a perturbing Hamiltonian. Also, since we assume $\chi\equiv N\max_{\,\ell
m\ell'm'}[U_{\ell m}^{\ell'm'}]/\Delta E\ll1$
we can consider all interlevel coupling terms as a perturbation as
well. This part of the Hamiltonian includes, first,  the zero-vorticity interlevel   and the vortex-antivortex interlevel transitions 
\begin{equation}
\hat{H}_{U_{00}^{1m}}=\sum_{j,m}\,U_{00}^{1m}
\!\left[\!\left(\!\bhat{j}{0}{0}{\dagger}\!\right)^{2}\!\bhat{j}{1}
{m}{}\bhat{j}{1}{-m}{}+\mathrm{h.c.}\right].\label{eq:PertH_U01}
\end{equation}
Secondly, it also includes the vortex-antivortex intralevel transitions 
\begin{equation}
\hat{H}_{U_{10}^{11}}=U_{10}^{11}\sum_{j,m}\!\left[
\!\left(\!\bhat{j}{1}{0}{\dagger}\!\right)^{2}\!\bhat{j}{1}{1}{}\bhat{j}{1}{-1}{
}+\mathrm{h.c.}\right].\label{eq:PertH_U11}
\end{equation}
In the following, we describe the perturbation theory for two cases: (i) the axisymmetric vortex and (ii) the transverse vortex, both discussed in the main text. For (i) only the tunneling processes given by Eq.~(\ref{eq:PertH_J}) are relevant. For  (ii),  the zero-vorticity interlevel transitions  in Eq.~(\ref{eq:PertH_U01}) and the vortex-antivortex intralevel transitions  in Eq.~(\ref{eq:PertH_U11}) are
also relevant. Both transitions are described in the main text.

\subsection*{Case (i): Axisymmetric Vortex} 

For $N$ even, the initial condition is the Fock vector
$|i\rangle$ with $n_{100}^{(i)}=N/2$ and $n_{211}^{(i)}=N/2$, and all other single particle modes unoccupied (see Fig~\ref{fig:Supl_Fig1}a). This Fock vector
is an eigenstate of the unperturbed Hamiltonian degenerate with
the Fock vector $|j\rangle$ with $n_{200}^{(j)}=N/2$ and $n_{111}^{(j)}=N/2$. For $N$ odd, one can consider  an extra atom in the ground mode [$n_{100}^{(i)}=(N+1)/2$ and $n_{211}^{(i)}=(N-1)/2$] or in the excited one [$n_{100}^{(i)}=(N-1)/2$ and $n_{211}^{(i)}=(N+1)/2$]. For the axisymmetric vortex case, the perturbing Hamiltonians Eqs.~(\ref{eq:PertH_U01})
and (\ref{eq:PertH_U11}) do not play any role, because the  matrix elements corresponding to vectors $|i\rangle$ and $|j\rangle$ are zero.
Then, the degenerate perturbation theory depends only on  the tunneling processes described by Eq.~(\ref{eq:PertH_J}),
and particularly on $\bhat{j}{0}{0}{\dagger}\bhat{j'}{0}{0}{}$ and
$\bhat{j}{1}{1}{\dagger}\bhat{j'}{1}{1}{}$. For $N$ even, the non-zero matrix element is obtained when both operators are applied $N/2$ times, thus giving a numerator of the corrections
to the eigenenergies proportional to $(N/2!)^2J_{00}^{N/2}J_{11}^{N/2}$. For $N$ odd this numerator is $[(N+1)/2!][(N-1)/2!]J_{00}^{(N\pm1)/2}J_{11}^{(N\mp1)/2}$, where the upper (lower) sign applies  if the extra atom is in the ground (excited) mode. For obtaining the denominator one has to consider all possible different
orders in which these two operators can be applied, and the difference
between the energies of the corresponding Fock vectors, which are the
eigenstates of the unperturbed Hamiltonian. For example, for $N=4$ the correction is
\[
\Delta\epsilon_{\pm1}=\frac{4J_{00}^{2}J_{11}^{2}}{\left(U_{00}+4U_{11}^{00}\right)8U_{00}^{11}\left(U_{11}+4U_{11}^{00}\right)}.
\]
For bigger values of $N$ the expressions become large but can be
shortcut as
\begin{eqnarray}
\Delta\varepsilon_{\pm1} & = &
2(N/2!)^{2}J_{00}^{N/2}J_{11}^{N/2}\tilde{U}_a\,,\label{eq:splitting_De1}
\end{eqnarray}
with $\tilde{U}_a$ defined in Eq.~(\ref{Eq:splittingN/2}) in the main text.
The corresponding eigenstates
are $\left(|i\rangle\pm|j\rangle\right)/\sqrt{2}$. For $N$ odd the expression for $U_a(n,p)$ has to  be adjusted to
\begin{align}
\label{Eq:splittingN/2_odd}
&U_a^\mathrm{odd}(n,p)\!=\!(-1)^{(n+p)}\left\{U_{00} 2n[(N\pm1)/2\!-\!n] \right.\nonumber\\
&+U_{11} 2p[(N\mp1)/2\!-\!p]\nonumber\\
&- \left.4U_{00}^{11}\left[(N\pm1)n/2+(N\mp1)p/2\!-\!2np\right]\right\}^{-1}\,,\nonumber
\end{align}
with  $(N+1)/2-1\le\sum_{j=1}^{N-1}i_{j}\le (N+1)/2$. The upper (lower) sign applies if the extra atom is in the ground (excited) modes.

\subsection*{Case (ii): Transverse Vortex} 

The initial condition is the Fock vector
$|i\rangle$ with $n_{100}^{(i)}=N/2$ and $n_{210}^{(i)}=N/2$, depicted
in Fig.~\ref{fig:Supl_Fig1} (b). We assume $N/2$ even in this case. This Fock vector is an eigenstate of the unperturbed Hamiltonian degenerate with the Fock vector
$|j\rangle$ with $n_{200}^{(j)}=N/2$ and $n_{110}^{(j)}=N/2$. Now, there are two different processes
in the perturbing Hamiltonian that give non-zero contributions to the perturbation. One is,
as in  case (i), the tunneling terms
$\bhat{j}{0}{0}{\dagger}\bhat{j'}{0}{0}{}$
and $\bhat{j}{1}{0}{\dagger}\bhat{j'}{1}{0}{}$, which leads to an
expression of the tunneling analogous to the previous one
\begin{eqnarray}
\Delta\varepsilon_{0} & = &
2(N/2!)^{2}J_{00}^{N/2}J_{10}^{N/2}\tilde{U}_a\,,\label{eq:splitting_De0}
\end{eqnarray}
with $\tilde{U}_a$ defined as in Eq.~(\ref{Eq:splittingN/2}) in the main text.
The other process is due to the zero-vorticity interlevel transitions 
$\left(\!\bhat{j}{0}{0}{\dagger}\!\right)^{2}\!\bhat{j}{1}{0}{}\bhat{j}{1}{0}{}
+\mathrm{h.c.}$
from the perturbing Hamiltonian (\ref{eq:PertH_U01}).   Since this process has to be applied in both wells, this leads to a numerator proportional
to $(N/2)!\left(U_{00}^{10}\right)^{N/2}$.  Again, to obtain the denominator
one has to consider all possible orders of applying this operator
in each well. For $N=4$ this gives the splitting
\[
\Delta\epsilon_{0}'=8\frac{\left(U_{00}^{10}\right)^{2}}{U_{00}-U_{11}-\Delta E}.
\]
For higher values of $N$ it is convenient to obtain a more compact expression of
the splitting
\begin{align}
\Delta\varepsilon_{0}'=2(N/2)\left(N/2!\right)^{2}\left(U_{00}^{10}\right)^{N/2}
\tilde{U}_t\,,\label{eq:splitting_De0p}
\end{align}
with $\tilde{U}_t$ defined as in Eq.~(\ref{eq:splittingN2t}) in the main text.
 Depending on the values of $J_{00}$, $J_{10}$, and $U_{00}^{10}/\Delta E$ the coupling can be dominated by tunneling or interactions.  If $\zeta_{00}$ is
bigger than $\chi$, expression~(\ref{eq:splitting_De0}) holds for the splitting while in the other case, expression~(\ref{eq:splitting_De0p})  holds; if $\zeta_{00}\simeq \chi$, the perturbation theory becomes more complicated, and we omit the expressions for brevity. Finally, the eigenstates are a combination of $|i\rangle$ and $|j\rangle$
with two other vectors $|k\rangle$ and $|l\rangle$, where $n_{100}^{(k)}=N/2$,
$n_{210}^{(k)}=N/2-2$, and $n_{21\pm1}^{(k)}=1$; and with $n_{200}^{(l)}=N/2$,
$n_{110}^{(l)}=N/2-2$, and $n_{11\pm1}^{(l)}=1$. Vector $|k\rangle$
is represented in Fig. \ref{fig:Supl_Fig1} (c). The coupling to these two vectors
is a consequence of the presence of term~(\ref{eq:PertH_U11}) in
the perturbing Hamiltonian, which we termed as the vortex-antivortex intralevel transitions  in the main text. This coupling is given by
\begin{align}
c_{U}=\frac{\sqrt{N/2}\sqrt{N/2-1}\,U_{10}^{11}}{U_{10}(2N-6)-2U_{11}-4(N/2-2)U_
{10}^{11}}.
\end{align}
Then, the eigenstates can be written as $\psi_{\pm}=\alpha|i\rangle\pm\alpha|j\rangle+\beta|k\rangle\pm\beta|l\rangle$
with $\alpha=1/\sqrt{2+2c_{U}^{2}}$ and $\beta=dc_{U}$.

Finally, for $N/2$ or $N$ odd , Eq.~(\ref{eq:splitting_De0p}) does not hold, because the zero-vorticity interlevel transitions  create/annihilate atoms in pairs. Then, to couple vector $|i\rangle$ to vector $|j\rangle$, it is necessary that at least one atom tunnels through the barrier.  If  $\chi$ dominates over $\zeta_{00}$, the tunneling energies $J_{00}$ and $J_{11}$ would be very small, which makes  $\Delta\varepsilon_{0}$ much smaller for the initial states with an odd number of atoms, as numerically shown in the main text.


\begin{thebibliography}{100}


\bibitem{1998ImadaRMP} M. Imada, A. Fujimori, and Y. Tokura, Rev. Mod. Phys. {\bf 70}, 1039 (1998); Y. Tokura and N. Nagaosa, Science {\bf 288}, 462 (2000)


\bibitem{2013PinheiroPRL} F. Pinheiro, G.~M. Bruun, J.-P. Martikainen,  J. Larson,  Phys. Rev. Lett. 111, 205302 (2013);  F. Pinheiro arXiv:1410.7828 (2014)

\bibitem{2011WirthNatPhys}  T. M\"uller,   S. F\"olling, A.  Widera, and I. Bloch, Phys. Rev. Lett. {\bf 99}, 200405 (2007);  G. Wirth, M. \"Olschl\"ager, and 	
 A. Hemmerich, Nature Phys. {\bf 7}, 147 (2011)

\bibitem{2011LewensteinNatPhys} M. Lewenstein and W. V. Liu, Nat. Phys. {\bf 7}, 101 (2011); A. Isacsson and S. M. Girvin, Phys. Rev. A {\bf  72}, 053604 (2005); V. W. Scarola and S. Das Sarma, Phys. Rev. Lett. {\bf 95}, 033003 (2005); W. V. Liu and C. Wu, Phys. Rev. A {\bf 74}, 013607 (2006); C. Xu and M.~P.~A. Fisher, Phys. Rev. B {\bf 75}, 104428 (2007);
A. Collin, J. Larson, and J.-P. Martikainen, Phys. Rev. A {\bf 81} 023605 (2010); N.~Y. Kim,
K. Kusudo, C. Wu, N. Masumoto, A. L\"offler, S. H\"ofling, N. Kumada, L. Worschech, A. Forchel, and Y. Yamamoto
Nature Physics {\bf 7}, 681 (2011); F.  H{\'e}bert,  Z. Cai, V.~G. Rousseau, C. Wu,  R.~T. Scalettar, and G.~G. Batrouni, Phys. Rev. B 87, 224505 (2013); F. Pinheiro, J.-P. Martikainen, and J. Larson, Phys. Rev. A 85, 033638 (2012)


\bibitem{footnote1}  Also called NOON, GHZ, or Schr\"odinger Cat states. 

\bibitem{2010PichlerPRA}  D.~A.~R. Dalvit, J. Dziarmaga, and W.H. Zurek Phys. Rev. A {\bf 62} 013607 (2000); H. Pichler, A. J. Daley, and P. Zoller, Phys. Rev. A {\bf 82} 063605 (2010)

\bibitem{Milburn1997PRA}  G.~J. Milburn,  J. Corney, E. M. Wright, and D. F. Walls, Phys. Rev. A {\bf 55} 4318 (1997);  A. Smerzi, S. Fantoni, S. Giovanazzi, and S. R. Shenoy, Phys. Rev. Lett. {\bf  79}, 4950 (1997);

\bibitem{2005AlbiezPRL} M. Albiez, R. Gati, J. F\"olling, S. Hunsmann, M. Cristiani, and M.K. Oberthaler, Phys. Rev. Lett. {\bf 95}, 010402 (2005); S. Levy,  E. Lahoud, I. Shomroni, and J. Steinhauer, Nature {\bf 449} 579 (2007)

\bibitem{2009SalgueiroPRA} J.~R. Salgueiro, M. Zacar\'es, H. Michinel, and A. Ferrando, Phys. Rev. A {\bf 79} 033625 (2009); O. Fialko, A. S. Bradley, and J. Brand, Phys. Rev. Lett. {\bf 108}, 015301 (2012); G. Watanabe and C.~J. Pethick Phys. Rev. A {\bf 76}, 021605 (2007)



\bibitem{2008AlonPRA} D. Masiello, S. B. McKagan, and W. P. Reinhardt, Phys. Rev. A {\bf 72}, 063624 (2005); A.~I. Streltsov, O.E. Alon, and L.S. Cederbaum, Phys. Rev. A {\bf 73}, 063626 (2006); K. Sakmann, A.~I. Streltsov, O.E. Alon, and L.S. Cederbaum, Phys. Rev. Lett.  {\bf 103}, 220601 (2009).





\bibitem{2010CarrEPL} L.~D. Carr, D. Dounas-Frazer, and  M.~A. Garcia-March, Europhys. Lett.  {\bf 90}, 10005 (2010);


\bibitem{2011GarciaMarchPRA}  M.~A. Garcia-March, D. R. Dounas-Frazer, and L.~D. Carr,  Phys. Rev. A {\bf 83}  043612 (2011)

\bibitem{2012GarciaMarchFiP} M.A. Garcia-March, D. R. Dounas-Frazer, and L.~D. Carr, Front. Phys. \textbf{7} 131 (2012)

\bibitem{2012GarciaMarch}  M.~A. Garcia-March and L.~D. Carr. In preparation

\bibitem{2000MadisonPRL} K.~W. Madison, F. Chevy, W. Wohlleben, and J. Dalibard, Phys. Rev. Lett. {\bf 84}, 806 (2000); J.~R. Abo-Shaer, 
C. Raman, J. M. Vogels, W. Ketterle,  Science {\bf 292}, 476 (2001); A. E. Leanhardt,  A. G\"orlitz, A. P. Chikkatur, D. Kielpinski, Y. Shin, D. E. Pritchard, and W. Ketterle, Phys. Rev. Lett. {\bf 89}, 190403 (2002)

\bibitem{2011BuckerNatPhys} 
R. B\"ucker, J. Grond, St. Manz, T. Berrada, T. Betz, C. Koller, U. Hohenester, T. Schumm, A. Perrin, and J. Schmiedmayer
Nat. Physics     {\bf 7},   608 (2011); R. B\"ucker, T. Berrada, S. van Frank, J.-F. Schaff, T. Schumm, J. Schmiedmayer, G. J\"ager, J. Grond, U. Hohenester
J. Phys. B: At. Mol. Opt. Phys. {\bf 46} 104012 (2013); S. van Frank, A. Negretti, T. Berrada, R. B\"ucker, S. Montangero, J.-F. Schaff, T. Schumm, T. Calarco, and J. Schmiedmayer, Nat. Communications    {\bf  5},  4009 (2014)

\bibitem{footnote2} For $N=8$ the matrices are of dimension $\Omega=6435$, and we used the MATLAB function {\it eig} to compute the whole spectra. In an Intel Core i7-2700 3.50GHz, 32Gb RAM, this takes approximately 4 minutes.


\bibitem{2010TsatsosPRA} M.~C. Tsatsos, A.~I. Streltsov, O.~E. Alon, and L.~S.
Cederbaum  Phys. Rev. A {\bf 82}, 033613 (2010)
 
\end{thebibliography}
\end{document}